# Validity of the Néel-Arrhenius model for highly anisotropic $Co_xFe_{3-x}O_4$ nanoparticles

T.E. Torres[1,2,3,†], E. Lima Jr.[4], A.Mayoral[1,3], A.Ibarra[1,3], C. Marquina[2,5] M. R. Ibarra[1,2,3] and G. F. Goya[1,2]

[1] *Instituto de Nanociencia de Aragón (INA), Universidad de Zaragoza, Zaragoza, Spain.*
[2] *Departamento de Física de la Materia Condensada, Facultad de Ciencias, Universidad de Zaragoza, Zaragoza, Spain.*
[3] *Laboratorio de Microscopias Avanzadas (LMA), Universidad de Zaragoza, Zaragoza, Spain.*
[4] *División Resonancias Magnéticas, Centro Atómico Bariloche/CONICET, S. C. Bariloche 8400, Argentina.*
[5] *Instituto de Ciencia de Materiales de Aragón (ICMA), CSIC - Universidad de Zaragoza, Zaragoza, Spain.*

**Abstract.**

We report a systematic study on the structural and magnetic properties of $Co_xFe_{3-x}O_4$ magnetic nanoparticles with sizes between 5 to 25 nm, prepared by thermal decomposition of $Fe(acac)_3$ and $Co(acac)_2$. The large magneto-crystalline anisotropy of the synthesized particles resulted in high blocking temperatures (42K< $T_B$< 345K for 5 < d < 13 nm) and large coercive fields ($H_C \approx$ 1600 kA/m for T = 5 K). The smallest particles (<d>=5 nm) revealed the existence of a magnetically hard, spin-disordered surface. The thermal dependence of static and dynamic magnetic properties of the whole series of samples could be explained within the Neel–Arrhenius relaxation framework by including the thermal dependence of the magnetocrystalline anisotropy constant $K_1(T)$, without the need of ad-hoc corrections. This approach, using the empirical Brükhatov-Kirensky relation, provided $K_1(0)$ values very similar to the bulk material from either static or dynamic magnetic measurements, as well as realistic values for the response times ($\tau_0 \approx 10^{-10}$s). Deviations from the bulk anisotropy values found for the smallest particles could be qualitatively explained based on Zener's relation between $K_1(T)$ and M(T).

[†] *Corresponding author: teo@unizar.es*

# I. INTRODUCTION

Ferrites are spinel oxides with formula $MFe_2O_4$ (M = 3d transition metal) with cubic crystal structure and a multiplicity of complex magnetic configurations arising from the diverse interactions between the M and Fe magnetic ions. When M = $Co^{2+}$, the resulting cobalt ferrite ($CoFe_2O_4$) has distinctive magnetic properties due to its large first order magnetocrystalline anisotropy constant ($K_1$= $2x10^5$ $J/m^3$), which is about an order of magnitude greater than any other spinel oxide.[1] Together with its chemical stability, this property make $CoFe_2O_4$ magnetic nanoparticles (MNPs) a fundamental material for magnetic recording applications and ferrofluids.[2] Considerable efforts have been made to obtain homogenous and stable water-based nanofluids through different synthesis routes such as hydrothermal, coprecipitation, microemulsion, forced hydrolysis, and reduction-oxidation methods.[2] In particular, the thermal decomposition of organometallic precursors in a boiling solution of organic solvents has been successfully used to produce MNPs with narrow size dispersion,[3,4] and thus they are being increasingly exploited in those applications with critical specifications about size dispersion of the MNPs.[5]

The ferrimagnetic order in $CoFe_2O_4$ results from the competing super-exchange interactions between the two magnetic sublattices of tetrahedral (A) and octahedral (B) sites in the structure. The $Fe^{+3}$ ions within the B sublattice are ferromagnetically ordered, as well as the $Co^{+2}$ ions within the A sublattice. On the other hand, the interactions between A and B spin sublattices are antiferromagnetic, resulting in an uncompensated net magnetic moment. The exchange energy in this material has been reported to be as large as $J_{AF}$ = -24 $k_B$.[6] It is well known that the relation between the anisotropy and exchange energies determines the critical size ($D_{cr}$) for the single domain configuration. The existence of a critical diameter $D_{cr}$ of a (spherical) particle implies that below a certain diameter value d such that d < $D_{cr}$, the lowest free energy state is that of uniform magnetization, as proposed by Brown.[7] This critical value has been estimated [8,9] to be $D_{cr} = 5.1\sqrt{\frac{A}{\mu_0 M_S^2}}$, where $A$ is the exchange stiffness[10] and $M_S$ is the saturation magnetization of the material. Using $A$ = 15 $x10^{-12}$ J/m; $M_S$ = 425 A/m (bulk $CoFe_2O_4$)[11] and $\mu_0 = 4\pi x 10^{-7}$ H/m, a critical diameter $D_{cr}$ = 40,7 nm is obtained. Accordingly, reported values of the single domain critical size for $CoFe_2O_4$ are between 30 and 70 nm.[12] As a consequence of the large magnetic anisotropy, single domain particles of $CoFe_2O_4$ of a few-nanometer size can retain the blocked regime up to room temperature. This particularity allows observing the thermal evolution of some magnetic parameters of MNPs such as saturation magnetization and coercivity of the blocked state in a wide range of temperatures before the superparamagnetic transition wipes out this information.

The energy E of an assembly of uniaxial particles with their easy axes parallel to the z axis under an external applied field is usually described (at T=0) by:

$$E(V) = K_{eff}V \sin^2\theta + HM_S V \cos\theta \qquad (1)$$

where θ is the angle between field H and saturation magnetization $M_S$, V the particle volume and $K_{eff}$ is the effective magnetic anisotropy. Assuming the energy of a single particle given by equation (1), the unblocking process occurs through an energy barrier ΔE given by:

$$\Delta E = K_{eff} V \left(1 - \frac{HM_S}{2K_{eff}}\right)^2 \qquad (2)$$

At a fixed temperature T the reversal of the magnetic moment occurs through the energy barrier given by equation (2). This thermally-activated process is described by the Néel-Arrhenius model, which gives a simple expression for the relaxation time $\tau = \tau_0 e^{K_{eff}V/kT}$. Taking τ = $10^2$ s for the measuring time window and $\tau_0 = 10^{-9}$ s we get $K_{eff}V = 25 k_B T$ the coercive field $H_C$ (T) can be expressed as:

$$\boldsymbol{H_C(T) = \frac{2K_{eff}}{M_S}\left[1 - \left(\frac{25k_B T}{K_{eff}V}\right)^{1/2}\right]} \qquad (3)$$

This is the well-known $H_C$ vs. $T^{1/2}$ relation often used for fitting the temperature evolution of the coercive field in the blocked state, i.e., at low temperatures. It is worth to note here that the thermal dependence of $K_{eff}$ in equation (3) is neglected, although previous studies of bulk spinel oxides have reported large variations of the anisotropy below room temperature.[13]

In this work, we report a systematic study on the magnetic properties in a series of Co ferrite magnetic nanoparticles within 5 and 25 nm. An exhaustive study by high resolution electron transmission microscopy (HRTEM) techniques has been performed in order to explore the influence of MNPs size and shape on the observed magnetocrystalline anisotropy,[14] with a precise observation of the crystallographic structure with atomic resolution. The chemical composition at the single-particle level was performed to assess the levels of stoichiometric homogeneity of samples. Systematic measurements of magnetization, coercive field and magnetic anisotropy were performed for increasing particle size to study the temperature evolution of the magnetic parameters in the blocked regime. The validity of the Neel-Arrhenius law for explaining the temperature dependence of the relaxation time has been re-gained by taking into account the variation of the anisotropy constant with the temperature.

## II. EXPERIMENTAL

$Co_xFe_{3-x}O_4$ nanoparticles of different sizes were prepared by thermal decomposition [3] of iron acetylacetonate Fe(acac)$_3$ and cobalt acetylacetonate Co(acac)$_2$ as precursors[4]. Different solvents (phenyl ether, benzylether, 1-octadecene, and trioctylamine) with increasing boiling temperatures were used in order to control the final particle size. For a standard preparation, 10.4 mmol of Fe(*acac*)$_3$ and 5.2 mmol of Co(*acac*)$_2$ were dissolved in 52 mmol of Oleic acid (OA), 65.4 mmol of Oleylamine, 86.5 mmol of 1,2 Octanediol and 150 ml of the chosen solvent. Then, the mixture was heated up to the stabilization temperature $T_{St}$ (200 °C in this case) under mechanical stirring under a flow of nitrogen gas for the nucleation step. This temperature was kept constant for 120 minutes, and then the solution was heated to the boiling temperature of the solvent (260-330°C), that is the final synthesis temperature, $T_{FSt}$, in nitrogen atmosphere. After waiting a few minutes (depending on the sample) at this temperature the solution was cooled down to room temperature. The resulting $Co_xFe_{3-x}O_4$ MNPs were washed three times with ethanol, and then magnetically-assisted precipitated until the supernatant solution became clear. Afterwards, the final product, composed by ferrite nanoparticles coated with a layer of oleic acid, was re-dispersed in hexane.

The samples were labeled as AVXX, where the number XX represents the average particle diameter (in nanometers) obtained from the core distributions observed in TEM images (see below). The resulting samples showed average particle diameters ranging from 5 to 25 nm. Details of the ether/alkenes used as solvents in each case, together with the stabilization ($T_{St}$) and final synthesis ($T_{FSt}$) temperatures used in each synthesis are given in Table SI of the supplemental material. In the case of sample AV11, the only sample synthesized in trioctylamine, the temperature was carefully raised for 10 minutes, from 320 °C up to $T_{FSt}$=330 °C. Once $T_{FSt}$ was reached the sample was immediately cooled down. It is also worth to mention that samples AV16 and AV18 were obtained from the same dispersion of nanoparticles by magnetically-assisted precipitation: sample AV18 was collected as the precipitated MNPs after applying a ferrite permanent magnet for 10 s to the as synthesized colloid and re-dispersing this precipitate in hexane. The supernatant resulting from this separation was precipitated a second time applying the magnet for 5 minutes, and re-dispersed in hexane. This latter sample was labeled as AV16. Sample AV25 was grown using the heterogeneous method [3] starting from already existing MNPs (sample AV13) as seeds, and mixing 80 mg with the same molar concentration of reactants.

The morphology and stoichiometry of the MNPs were studied by Transmission and Scanning Electron Microscopies (TEM and SEM, respectively). TEM images were obtained using a thermo-ionic $LaB_6$ 200 kV Tecnai T20 microscope operating at an accelerating voltage of 200 kV. STEM–HAADF (Scanning Transmission Electron Microscopy using a High Angle Annular Dark Field detector) images were acquired using a XFEG TITAN 60–300 kV, operated at 300 kV, equipped with monochromator and with a CEOS hexapole aberration corrector for the electron probe. TEM specimens were prepared by placing a drop of a hexane solution containing the MNPs onto a holey carbon coated copper micro-grid. The mean particle size <d> and size distribution were evaluated by measuring about 150-500 particles found in arbitrarily chosen areas of enlarged micrographs of different regions of the micro-grid. SEM measurements were carried out in a FEI INSPECT F with INCA PentaFETx3 system operating at 20 keV. The ratio between iron and cobalt content was determined through Energy-Dispersive X-ray spectroscopy (EDX) performed on a macroscopic zone of a powder sample (about 10000 μm$^2$) in SEM analyses, and on a small area (about 1000 nm$^2$) containing many particles as well as on single particles using the TEM.

The total iron concentration was determined from UV/Vis spectroscopy in a Varian Cary 50 Spectrophotometer operating at a fix wavelength of 478 nm. For the absorbance measurements, Potassium thiocyanate (KSCN) was used following the standard protocol described elsewhere.[15,16]

Magnetization measurements $M$ ($T$, $H$) and ac magnetic susceptibility measurements were performed on a MPMS-XL SQUID Quantum Design magnetometer. All measurements were performed on dried samples, after conditioning the dry powder inside plastic capsules. The temperature dependence of the magnetization was measured following zero-field-cooling (ZFC) and field cooling (FC) protocols, applying 7.9 kA/m, and the data were collected increasing the temperature from 5 to 400 K. The magnetization isotherms were measured between 5 and 400 K up to a maximum magnetic field of 3.96 MA/m. The susceptibility versus temperature was measured applying an excitation ac field of ~0.24 kA/m, at frequencies from 0.1 to 10$^3$ Hz, under zero external dc magnetic field.

## III. RESULTS AND DISCUSSION

### A. Particle morphology and composition analyses

The analysis of the TEM images (Fig. 1) showed that for each particular synthesis, the MNPs obtained can be considered as uniform in size. The statistical analysis of the MNPs size distribution done by fitting the respective size-histograms to a Gaussian distribution yielded mean diameters ranging from <d> = 5 to 25 nm and standard deviations size distribution widths σ between 0.7 and 3 nm (see Table I).

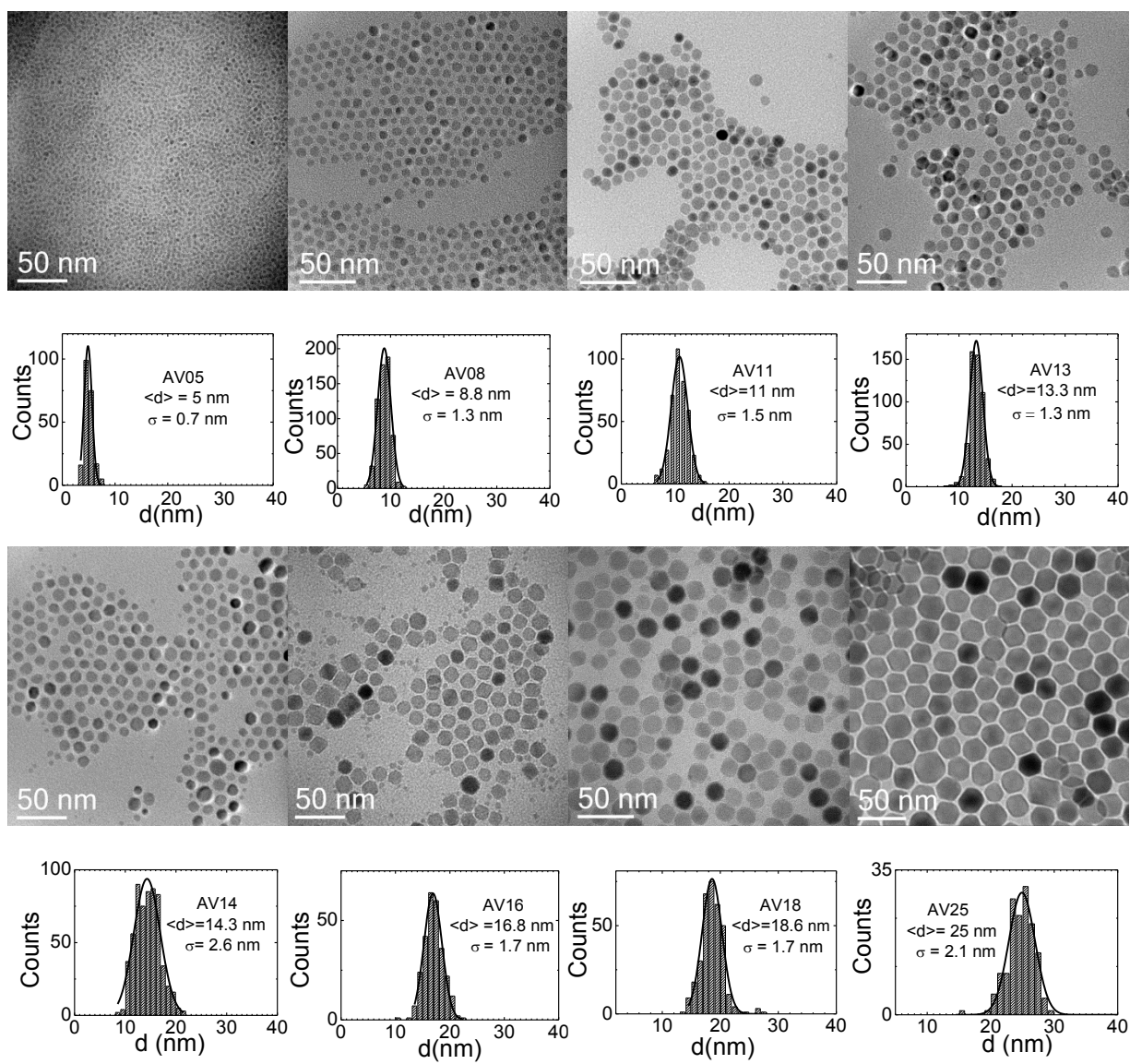

**Figure 1**. TEM images of CoFe$_2$O$_4$ nanoparticles for the AVXX series. The corresponding size histograms are shown below each image, together with the Gaussian fit, (solid lines) and the obtained mean size <d> and distribution width ($\sigma$). All micrographs were taken at the same magnification.

As previously reported for this synthesis route, the final average particle size reflected the influence of both the boiling point of the solvent and boiling time.[3, 17] Specifically, a systematic increase of the average particle size <d> for increasing boiling temperature of the solvent was observed. In the case of sample AV11 the final size is a combination of the higher boiling temperature and a shorter time at $T_{Fst}$ (10 minutes, see Table SI of supplemental material).[18] Regarding the MNPs morphology, the analysis of HRTEM images showed that for <d> ≤ 13 nm a noticeable population of rounded-shaped particles were present, whereas the largest ones showed a more faceted structure (see

Figs 1 and 2 and Fig. S1 in supplementary material). It has been proposed that the different morphologies are related to the rate of the temperature increase from the stabilization temperature ($T_{St}$) to the final synthesis temperature ($T_{FSt}$), and to the total reaction time at $T_{FSt}$.[18] Assuming the thermal decomposition as an autocatalytic reaction[4] it is expected that the concentration of precursor in the solution, which is inversely proportional to the volume of the particle, has a time dependence described by the logistic equation [19]. For the samples prepared with $T_{FSt}$ = 320 and 330 °C a linear dependence of <d> with the time of solution at $T_{FSt}$ has been observed, suggesting that the reaction is in an intermediate time regime without the complete consumption of the precursor. In addition, the composition analyses presented later on also show a time dependence of the composition on $T_{FSt}$, for $T_{FSt}$ = 320 and 330 °C, indicating that the chemical kinetics of Co and Fe incorporation onto the particle are different. For lower $T_{FSt}$ temperatures, we observe smaller values of <d>.

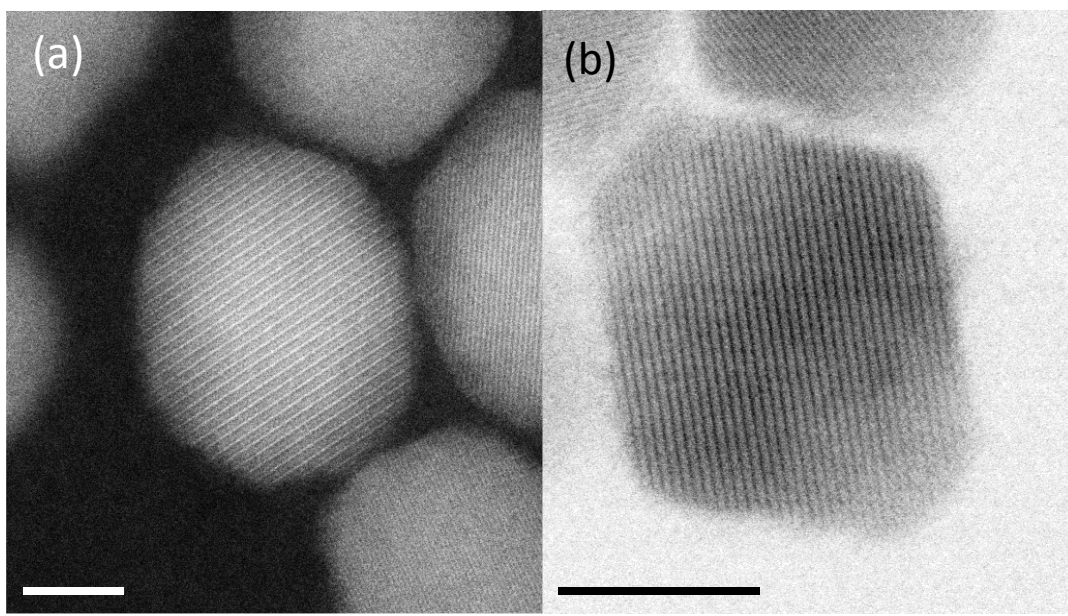

**Figure 2.** (a) $C_s$-STEM–HAADF image of sample AV13 and (b) $C_s$-STEM–BF image of sample AV18. Spherical and faceted morphologies are observed.

The $C_S$-corrected STEM-HAADF analysis at atomic resolution revealed that all the synthesized nanoparticles crystallized in the spinel structure with *Fd-3m* space group and unit cell parameter a = 8.394 Å. The data showed no evidence of distortions, crystal defects or any preferential orientation of the nanoparticles. As an example, Fig. 3(a) shows a high resolution $C_S$-STEM-HAADF image of a particle of sample AV13.

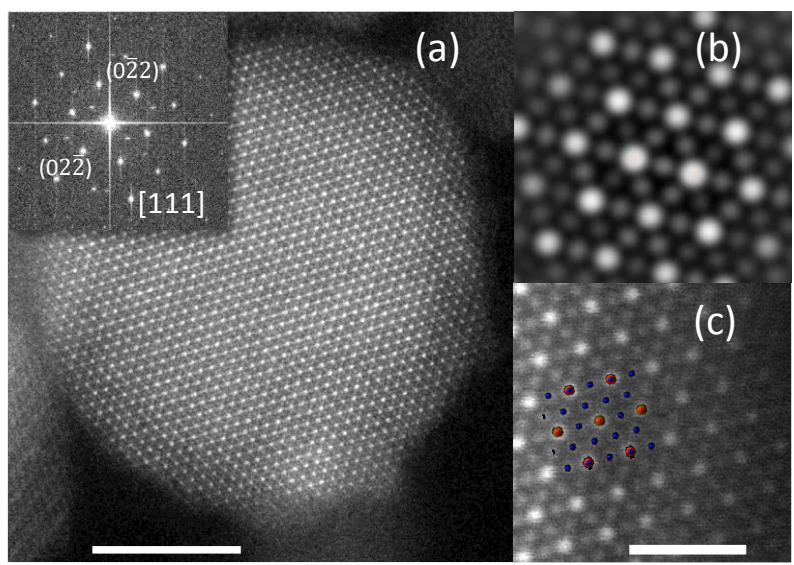

**Figure 3.** (a) High-resolution $C_s$ STEM–HAADF image of a particle of the AV13 sample with its correspondent FFT inset. (b) Simulated image (c) a magnified region displaying the atomic distribution with the model superimposed.

The inset corresponds to the Fast Fourier Transform (FFT) in the [111] zone axis, showing the spots corresponding to $(0\bar{2}2)$ and $(02\bar{2})$ planes. Fig. 3(b) shows a simulated image using the parameters of the $CoFe_2O_4$ structure from the *Crystallography Open Database* (St. sample card Nº 22-1086 of the JCPDS-International Centre for Diffraction Data®-ICDD®). Finally, Fig. 3(c) displays the superposition of the simulated and real crystal structures, showing the coincidence of both of them.

The relative abundance of cobalt and iron in the samples was obtained from EDX in SEM analyses, taking spectra in different zones of the sample. In SEM the electron beam spot has a diameter between 10-100 nm and the emitted X-rays are collected from an underlying sample volume of about 1-3 μm deep. Therefore the information of the atomic composition corresponds to a volume around 0.2 μm$^3$ and therefore these results reflect the 'macroscopic' average composition of the sample. As an example, a micrograph corresponding to sample AV14 (<d> = 14.3 nm) is shown in Fig. S2(a) of the supplementary material, indicating the squared-defined area for EDX–SEM sampling. The corresponding EDX spectrum from this area (Fig. S2(b) of the supplementary material) showed the peaks associated with the $K_\alpha$ and $L_\alpha$ edges of iron and cobalt atoms. A minimum of five areas within the sample holder were studied for each sample, and in all cases the results were coincident within the experimental error, supporting the macroscopically homogeneous nature of the samples. The results are summarized in Table I. A deviation from the stoichiometry (i.e., atomic ratio $\rho$=[Fe]/[Co]=2.0) can be noticed, showing an excess of iron in all the samples. The resulting composition of the $Co_xFe_{3-x}O_4$ MNPs extracted for these analysis yielded x values ranging from 0.90 (sample AV05) to 0.54 (for sample AV08).

Table I: Average particle diameter <d> with deviation σ, atomic Fe/Co ratio (ρ) obtained by EDX-SEM and EDX-TEM, and the resulting chemical composition $Co_xFe_{3-x}O_4$

| Sample | <d> (nm) | σ (nm) | ρ=[Fe]/[Co] EDX-SEM | ρ=[Fe]/[Co] EDX-TEM | $Co_xFe_{3-x}O_4$ |
|---|---|---|---|---|---|
| AV05 | 5.0 | 0.8 | 2.3 | 2.9 | $Co_{0.90}Fe_{2.10}O_4$ |
| AV08 | 8.8 | 1.3 | 4.5 | | $Co_{0.54}Fe_{2.46}O_4$ |
| AV11 | 11.0 | 1.6 | 2.9 | | $Co_{0.77}Fe_{2.23}O_4$ |
| AV13 | 13.3 | 1.3 | 3.4 | | $Co_{0.68}Fe_{2.32}O_4$ |
| AV14 | 14.3 | 2.6 | 3.5 | 4.6 | $Co_{0.67}Fe_{2.33}O_4$ |
| AV16 | 16.8 | 1.7 | 3.9 | | $Co_{0.61}Fe_{2.39}O_4$ |
| AV18 | 18.6 | 1.7 | 3.5 | | $Co_{0.66}Fe_{2.34}O_4$ |
| AV25 | 25.0 | 2.1 | 3.3 | 3.6 | $Co_{0.70}Fe_{2.30}O_4$ |

Analysis of the chemical composition was also performed through TEM at the single-particle level, by acquiring the EDX spectra of individual particles and small aggregates of MNPs for samples AV05, AV14 and AV25. Typical results obtained for sample AV25 (<d>= 25 nm) are displayed in Fig. 4. For all analyzed samples, the Fe:Co ratios derived from individual particles and from particle clusters coincide, as in the EDX-SEM analysis. The close values of both TEM and SEM analysis in each case (see Table I) indicate that the chemical composition of the MNPs is homogeneous throughout the samples and, more importantly, within individual particles. Clearly, this analysis of the homogeneous internal structure of single MNPs is performed only for a few selected MNPs. However, the consistency of these data from several particles has been verified in *all synthesized samples* and therefore, gives support to the statistical confidence of these results.

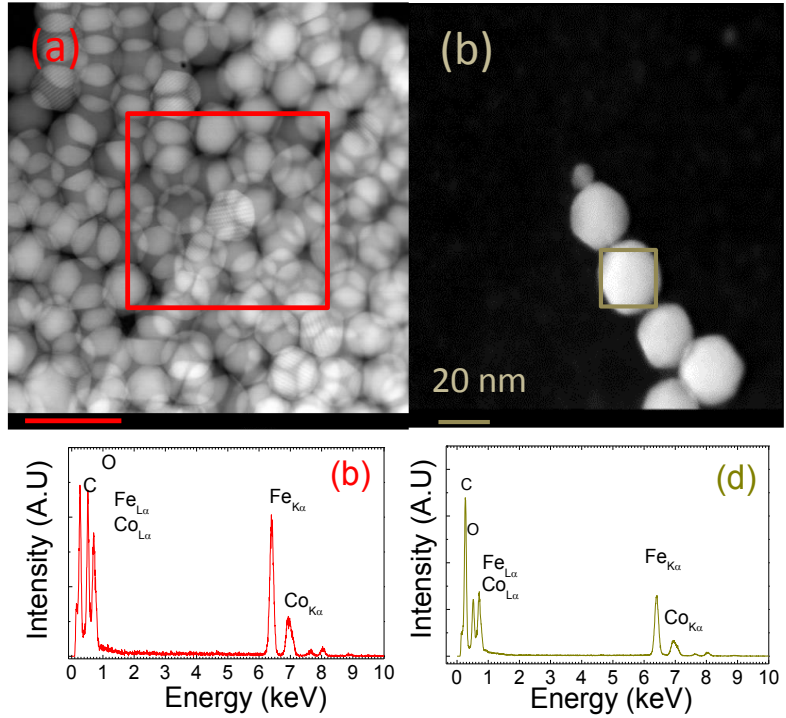

**Figure 4.** EDX-TEM carried out on sample AV25 (<d>= 25 nm). Spectra in (c) and (d) correspond to the nanoparticles in the area selected in (a) and to the particle selected in (b).

## B. Temperature dependence of the magnetization.

The main features of magnetization M(T) curves, taken in zero-field cooling and field-cooling (ZFC/FC) modes for all samples exhibited similar trends, as can be seen from Fig. 5(a). The blocking temperature distributions were obtained from the plot of $\left(\frac{1}{T}\right)\frac{d(M_{ZFC}-M_{FC})}{dT}$ vs. $T$ (Fig. 5b), and the mean blocking value <T$_B$> was extracted by fitting the experimental data with a Gaussian distribution. Large <T$_B$> values were obtained even for the smaller samples (T$_B$ = 42 K for particles with <d>= 5nm), reflecting the large magnetic anisotropy of CoFe$_2$O$_4$.[17,20,21,22,14] For those particles larger than 14 nm, the blocking temperatures were beyond the maximum of our experimental setup. It is interesting to note that the shift of <T$_B$> to higher temperatures with increasing particle size was not linear with particle volume V as expected from the functional definition of <T$_B$> in equation (3), i.e., $T_B = K_{eff}V/25k_B$. Instead, a nearly linear dependence on particle diameter <d> was observed.

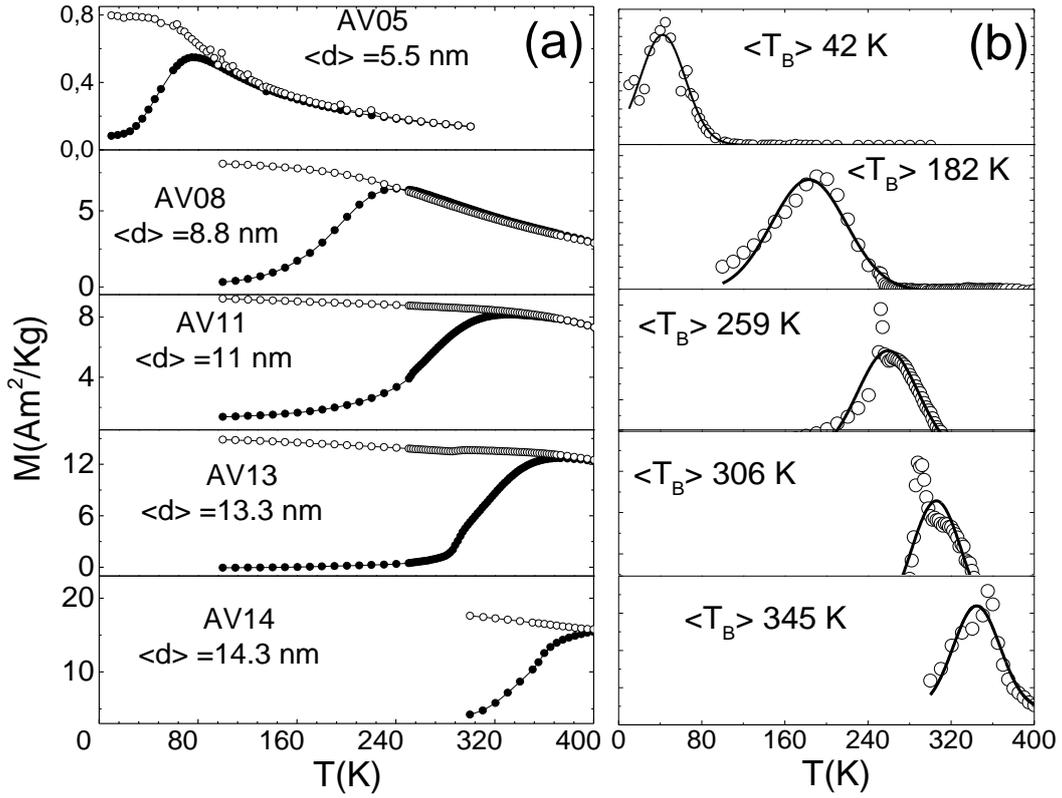

**Figure 5.** (a) M (T) data taken in zero-field-cooled (ZFC) and field-cooled (FC) modes for samples AV05 to AV14 (<d> between 5 to 14.3 nm). (b) Blocking temperature distributions fitted with a Gaussian function (solid line).

## C. Magnetic field dependence of the magnetization.

The magnetization of all samples was studied at temperatures from 5 K to 400 K, in applied field H up to 11.2 MA/m (14 T). For the M(H) performed at T = 400 K, the obtained coercive field values $H_C$ decreased with decreasing particle size (Table II) attaining zero for the samples with d < 13 nm, in agreement with the blocking temperatures observed from ZFC/FC curves. At T = 5 K, the hysteresis loops (figure 6) showed similar features for all samples, i.e., large coercive fields $H_C$ and saturation magnetization values $M_S$ around 60 Am$^2$/kg. The values of $M_S$ at 5 K collected in Table II are lower than the typical ($M_S$ = 80 Am$^2$/kg) found for bulk CoFe$_2$O$_4$.[11] This reduction of $M_S$ has been previously observed for small particles (1-10 nm) and thin films[23, 24] and it could be related to changes in the inversion degree of the spinel configuration. Indeed, there is no clear consensus about the inversion degree of cobalt ferrite in bulk and in nanostructured forms, probably because the relative occupancy of the A and B sites by Co and Fe seems to depend on sample preparation details. While neutron diffraction studies[25] have indicated that bulk $CoFe_2O_4$ has an inverted spinel configuration, latter Mossbauer and X-ray spectroscopy data[26, 27] indicated a partially inverted configuration, consistent with inversion degrees as high as $i=0.76$ in the formula $[Co_{1-i}Fe_i]^A[Co_iFe_{(2-i)}]^BO_4$.[28] A second explanation for the observed reduction in $M_S$ could be the existence of spin canting at the particle surface [11, 29] originated from competing interactions between A and B sublattices when a symmetry break and oxygen vacancies are produced at the particle surface. Monte Carlo simulations using different models[30,31] and approximations have shown that the reduction of $M_S$ is size dependent, and is related to the canted configuration of the spins at the surface.

Table II: Blocking temperature $<T_B>$, coercive field $H_C$ and saturation magnetization $M_S$ of $Co_xFe_{3-x}O_4$ samples with different average particle diameters, $<d>$.

| Sample | $<d>$ (nm) | $<T_B>$ (K) | $H_C$ (kA/m) 5 K | $H_C$ (kA/m) 400 K | $M_S$ (Am²/kg) 5 K | $M_S$ (Am²/kg) 300 K | $M_S$ (Am²/kg) 400 K |
|---|---|---|---|---|---|---|---|
| AV05 | 5 (0,8) | 42 | 390 | 0 | 30 | 24 | 17 |
| AV05* | | | 1060 | 0 | 31 | | 61 |
| AV08 | 8,8(1,3) | 182 | 1600 | 0 | 54 | 43 | 40 |
| AV11 | 11 (1,6) | 259 | 920 | 0 | 61 | 51 | 47 |
| AV13 | 13,3(1,3) | 306 | 1600 | 0 | 86 | 76 | 67 |
| AV14 | 14,3(2,6) | 345 | 1600 | 2 | 66 | 57 | 53 |
| AV16 | 16,8(1,7) | >400 | 1400 | 3 | 55 | 47 | 45 |
| AV18 | 18,6(2,1) | >400 | 1500 | 10 | 57 | 47 | 51 |
| AV25 | 25(4,1) | >400 | 1030 | 500 | 53 | 48 | 44 |

*Values of $M_S$ and $H_C$ for sample AV05 obtained from the high-field M(H) cycles (up to H = 11.4 MA/m) at 5 K and 400 K.

For all but AV05 and AV08 samples (i.e., the two smallest particle sizes), the magnetization was nearly saturated at $H = 2 \times 10^3$ kA/m. Samples AV05 showed a marked decrease in the magnitude of M, and no signs of saturation up to the highest field. We further investigate this behavior of sample AV05 through measuring the M(H) curves up to H = 11.2 MA/m at 400 K and at 5 K (see Figure 7). As expected for a minor loop, saturation was not reached even at this high field and the cycle remained open showing that the irreversibility field $H_{irr}$, defined as the field where the two branches of the hysteresis loop merge, was larger than our attainable maximum field.

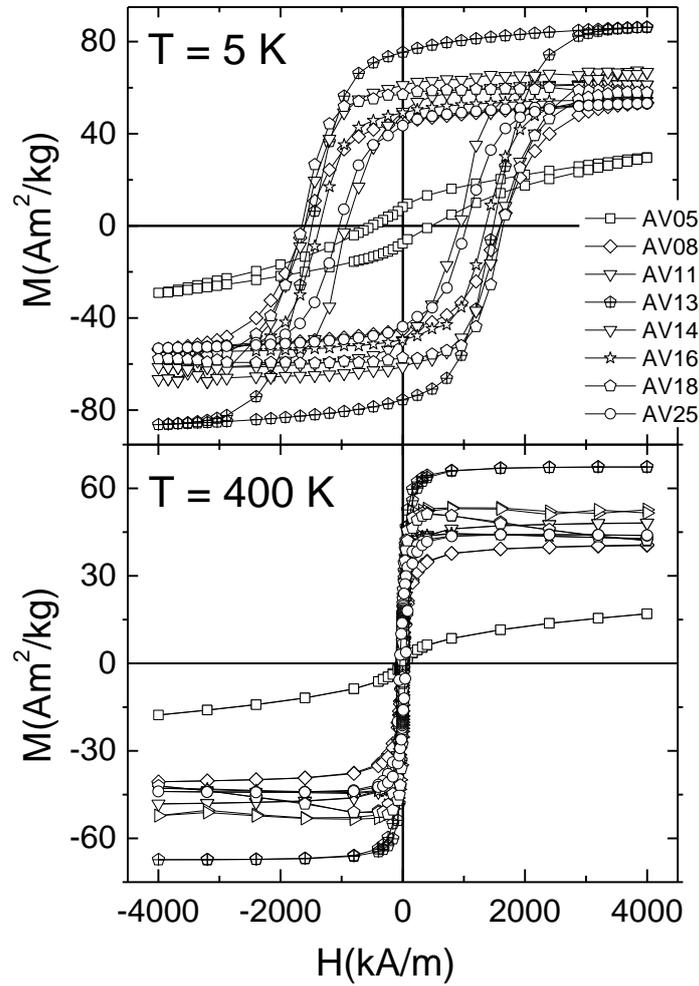

**Figure 6.** M (H) curves for all samples measured at (a) T = 400 K and (c) 5 K

The hypothesis of the surface spin canting that could explain the reduction of magnetization, also would originate the non-saturating behavior of the M(H) curves even at large applied fields, similarly to previous reports on small-sized ferrite nanoparticles. [21, 28, 32] This is likely to be the case in our samples AV05 and AV08, with a less pronounced effect in AV08 since surface effects are attenuated in particles with increasing volume.

For the rest of the samples, however, the decreasing surface/volume ratio would imply that surface spin canting cannot be a major cause for magnetization reduction. Additionally, for these samples the observed reduction in $M_S$ is not accompanied by the linear increase of the magnetization at high fields. On the contrary, the M (H) curves showed that the magnetic saturation is attained at moderate fields (H ≅ 2 MA/m), consistent with previous findings using polarization-analyzed small-angle neutron scattering experiments on Co-ferrite nanoparticles of 11 nm. [33,34] These results are in agreement with our observation of the concurrent low value of the saturation magnetization and the small fields required to reach $M_S$.[33,34] There is experimental evidence that the above mentioned spin canted structure extends over the whole particle volume, instead of forming a shell.[28] In moderate/high magnetic fields the measured magnetization is due to the net sum of spin components parallel to the applied field, and the reduction with respect to the bulk magnetization is due to the cancellation of the components perpendicular to the field, as the result of the competition between Zeeman and anisotropy energies. This might be the case of our nanoparticles with <d> ≥ 11 nm, being the particles with <d> = 13 nm, those in which the canting angle is lower (and therefore the magnetization is higher). However, local probe and/or neutron scattering experiments would be necessary to confirm this hypothesis.

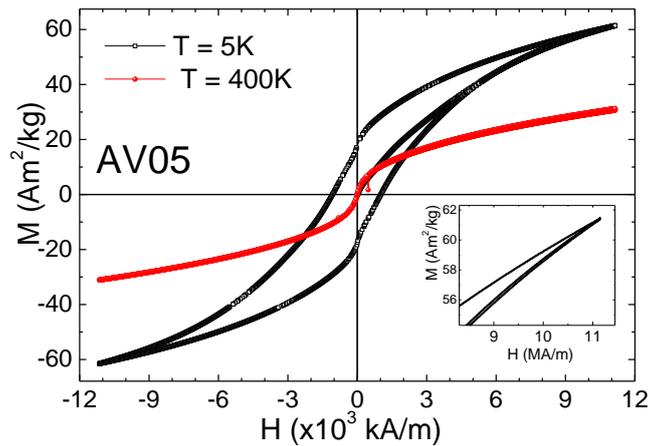

**Figure 7.** Magnetization hysteresis curves measured at 400 and 5 K for sample AV05 measurements taken until 11.2 MA/m (14T). The inset shows the high-field irreversibility from the T = 5 K data.

For AV05, the drastic reduction of magnetization observed in Figure 6 goes together with a clear non-saturating behavior up to H = 4 MA/m, also observed (although much less pronounced) for sample AV08. Additionally, irreversible behavior up to the largest fields (i.e., non-closure of the M(H) loops) could be observed for AV05 sample. These effects have been observed in many systems like $ZnFe_2O_4$ [29] and $CuFe_2O_4$ [35] ferrites and was first explained by J.M.D Coey [36] as originated from a spin-canted configuration of the surface spins due to broken symmetry at the surface and/or to oxygen-deficient stoichiometry. To get further understanding of this process, high-field measurements of sample AV05 were performed up to H ≤ 11 MA/m at both T = 5 K and 400 K (see Figure 7). The non-saturation observed in AV05 at the highest fields of ~11 MA/m implies anisotropy fields much larger than the expected from magnetocrystalline or shape anisotropy as sources of magnetic anisotropy, and suggests that spin canting (originated in exchange interactions) must be operative. In agreement with our results, previous reports by Respaud *et al.*[37] attributed the linear increase of M(H) up to fields of 28 MA/m observed in ultrasmall cobalt nanoparticles to the major influence of surface atoms as particle size decrease. Given the small particle size of AV05 samples, the increasing contribution from surface atoms to the overall magnetic moment is the more likely explanation for this M(H) behavior. The existence of a large number of broken exchange bonds at the surface of the particle, associated to the lack of neighboring atoms has been modeled by a shell of misaligned spins that surrounds a magnetically ordered core.[38]

The values of $H_C$ measured at T = 5 K and 400 K (see Table II) are in agreement with previously reported results in nanosized cobalt ferrite.[21, 39-41] The values observed at low temperature are within ≈1< $H_C$ < 1.6 MA/m. As the magnetization isotherm of sample AV05 corresponds likely to a minor loop, its corresponding small $H_C$ value cannot be compared with those of the rest of the series. The value observed for particles with <d>= 25 nm is in good agreement with the theoretical calculations performed by Kachkachi *et al.*[31] that predicted lower coercivity in faceted nanoparticles as compared to spherical ones, due to the higher symmetric coordination of surface atoms and lower amount of missing coordinating oxygen atoms. However, due to the synthesis protocol mentioned in Section II, a mixture of spherical and faceted particles cannot be discarded. These synthesis conditions might have also resulted in a distribution of Fe and Co atoms among A and B crystallographic sites different than the rest of the series. The change in site populations would lead to a different local anisotropy of $Co^{2+}$ ions, which could explain the observed lower value of $H_C$.

**D. Temperature dependence of the coercive field.**

We have studied the evolution of the coercive field, $H_C$, with the temperature by plotting the experimental $H_C(T,V)$ data for 5 ≤ T ≤ 400 K. The expected decrease of $H_C(T)$ for increasing temperature was observed in all samples, reaching the $H_C$=0 value at the corresponding superparamagnetic transition temperatures. The exact functional dependence of $H_C$ with temperature for single/domain magnetic nanoparticles in the blocked state has been discussed since decades ago. Within the simple Neel-Arrhenius model already presented in the introduction section, a $H_C \propto T^{1/2}$ is expected. However, equation (3) neglects the particle size dispersion existing in any real sample, which is an oversimplification in most cases.[42] Recent works have pointed out the difficulties of including the size distribution into a realistic model[43] because the measured $H_C$ is not a simple superposition of individual particle coercivities. An analytical expression for the dependence of $H_C(T)$ with T and particle size has been proposed,[44] obtaining a $T^{3/4}$ for the thermal dependence in a randomly-oriented ensemble of particles. The fact that this approximation was unable to fit our experimental data for any sample, together with the quite narrow size distributions observed in our samples (see Figure 1) suggest that deviations from the $T^{1/2}$ law for $H_C$ were not due to size distributions.

The departures observed from the $H_C(T)$ vs. $T^{1/2}$ graphs of our samples (see Figure S3 in the supplementary material) were increasingly marked for the larger particles, strongly suggesting that this feature was related with some neglected T-dependence of the magnetic parameters involved. As equation (3) assumes that the magnetocrystalline anisotropy is a temperature-independent parameter, the corresponding $H_C$ expression should be a valid approximation only for a narrow T-range where $K_1$ is not expected to vary substantially.[45, 46] This is the case for particles with low blocking temperatures, since only in the blocked state $H_C>0$ can be effectively measured. Indeed, a good $T^{1/2}$ fits have been reported for small and/or low-anisotropy MNPs (e.g., T< 50 K).[47,48, 49] However, this approximation fails completely for particles with large size and/or anisotropies like $CoFe_2O_4$, for which the blocked state may span a temperature range from 5 to 400 K. In such a wide temperature interval $K_1(T)$ can change markedly[50] and therefore, the $T^{1/2}$ dependence of $H_C$ is no longer valid. The importance of the temperature dependence of the anisotropy has been pointed out in previous works in relation to the thermal dependence of $H_C$ of metallic Fe, Co and Ni[8,51,52] nanoparticles, as well as in Co-containing ferrites.[53,54,55,56] However, an explicit thermal dependence of the magnetocrystalline anisotropy has not been so far included in the expression of $H_C(V,T)$, to the best of our knowledge.

The classical theory by Zener [57] on the effect of temperature on the magnetic anisotropy provides a relation between the magnetization M and $K_1$ of the form[58]

$$\frac{K_1(T)}{K_1(0)} = \left[\frac{M(T)}{M(0)}\right]^n \qquad (4)$$

with n = 10 for full correlation between adjacent spins and n=6 for incomplete correlation.[59] In cubic ferromagnetic crystals like spinel oxides, this relation is expected to hold for temperatures below $0.9T_C$, being $T_C$ the Curie temperature of the material. Based on these relationships Shenker[60] has demonstrated that for bulk cobalt ferrite $K_1(T)$ can be expressed by the empirical Brukhatov-Kirensky relation [60]

$$\boldsymbol{K_1(T) = K_1(0)\exp(-BT^2)} \qquad (5)$$

valid for the 20 K< T < 350 K temperature range, with $K_1(0) = 1.96 \times 10^6 \, J/m^3$ and $B = 1.9 \times 10^{-5} \, K^{-2}$. Incorporating this dependence into the $H_C(T)$ expression given by eq. (3) and considering that $K_{eff}$ as the first magnetocrystalline anisotropy constant $K_1$ we obtain:

$$H_C(T) = \frac{2K_1(0)\, e^{-BT^2}}{\mu_0 M_S}\left[1 - \left(\frac{25 k_B T}{V A e^{-BT^2}}\right)^{1/2}\right] \qquad (6)$$

As seen in Figure 8, this expression provides an excellent fit of the experimental data for a wide range of particle sizes and temperature, and makes clear that any attempt of describing the thermal evolution of any magnetic parameter depending on $K_{eff}$ over more than a few-degrees temperature range should consider the impact of $K_1(T)$.

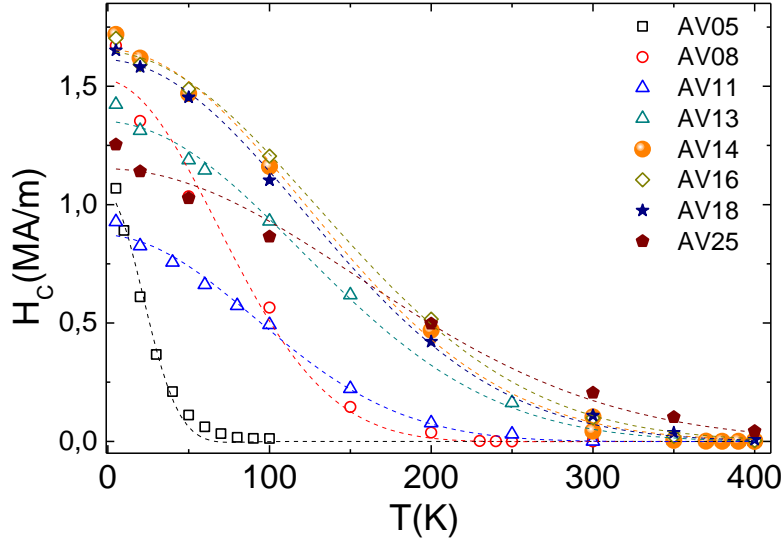

**Figure 8.** Temperature dependence of the coercive field $H_C$. The dashed lines are the corresponding fit using $H_C(T)$ given by eq. (6).

The values of $K_1(0)$ and B obtained from Figure 8 using eq.(6) are listed in Table III. They are in excellent agreement with previous experimental reports[53,61,56,62] and theoretical calculations[63,64] for this material. For those samples with <d> between 13- 25 nm the obtained $K_1(0)$ values spanned a narrow range 2.8-5.4x10$^6$ J/m$^3$, with a maximum difference of ≈60% from the bulk value in sample AV16.

The magnetocrystalline anisotropy of $CoFe_2O_4$ is due to the spin-orbit coupling, mainly from the contribution of the $Co^{+2}$ cations at the octahedral B sites. Therefore, changes in the occupancy factor of A and B sites usually reported in many spinel ferrites[65,66] could be expected to yield changes in $K_1$ values. The fact that the chemical composition of our nanoparticles is off-stoichiometric would have led us to expect this departure in the cobalt content to influence the magnetocrystalline anisotropy as well. Our data showed no major deviations from nominally stoichiometric bulk samples regarding magnetic anisotropy. For the smallest samples AV05 and AV08, an increase in both $K_1$ and B fitted parameters can be noticed. As the B parameter is related to the *n* exponent of Zener's relation, it seems plausible that the non-saturation behavior due to the spin canting will translate in large deviations of the M(T)/M(0) ratio, thus affecting the B parameter. Similar arguments could be applied to qualitatively explain the additional contribution to the anisotropy observed for $K_1(0)$ in AV05 and AV08 samples.

Table III. Parameters $K_1(0)$ and B obtained from a) fitting the $H_C(T)$ data using eq.(6); and b) Néel–Arrhenius model using the eq. (8). For the latter, the values of $\tau_0$ are also listed.

|        | Diameter (nm) | (a) K₁(0) (x10⁶ J/m³) | (a) B (x10⁻⁵ K⁻²) | (b) K₁(0) (x10⁶ J/m³) | (b) B (x10⁻⁵ K⁻²) | (b) τ₀ (x10⁻¹⁰ s) |
|--------|---------------|------------------------|--------------------|------------------------|--------------------|--------------------|
| AV05   | 5.0           | 7.3(2)                | 87(7)             | 0.41                  | 8.12              | 16.7              |
| AV08   | 8.8           | 5.6(1)                | 8.7(1)            | 0.59                  | 2.45              | 8.14              |
| AV11   | 11.0          | 2.77(4)               | 4.6(2)            | 2.12                  | 1.94              | 5.61              |
| AV13   | 13.3          | 2.86(2)               | 2.8(1)            | 3.61                  | 2.47              | 3.81              |
| AV14   | 14.3          | 4.60(4)               | 2.8(1)            |                        |                    |                    |
| AV16   | 16.8          | 5.41(5)               | 2.7(1)            |                        |                    |                    |
| AV18   | 18.6          | 5.28(5)               | 2.7(1)            |                        |                    |                    |
| AV25   | 25.0          | 3.79(9)               | 1.9(2)            |                        |                    |                    |
| Bulk*  |               | 1.96                  | 1.9               |                        |                    |                    |

* Values from Ref.[60]

### E. Temperature and frequency dependence of the AC magnetic susceptibility

In order to get a deeper insight into the effective magnetocrystalline anisotropy obtained from dc data, the magnetic dynamics of these nanoparticles was studied through the temperature dependence of $\chi'$ and $\chi''$ at fixed field amplitude and increasing frequency from 100 mHz to 1 kHz. Typically, both $\chi'(T)$ and $\chi''(T)$ components for all samples exhibited the peak at a temperature $T_P$ expected for a single-domain magnetic particle, which shifted towards higher T values with increasing frequency. Typical curves are shown in Figure 9 as examples for <d> = 8.8 and 11 nm (samples AV08 and AV11, respectively). The dynamic response of an ensemble of single-domain magnetic nanoparticles can be described by the thermally-assisted magnetic relaxation of a single-domain magnetic moment over the anisotropy energy barrier $E_a$.[47] The relaxation time $\tau$ associated to this process is given by a Neel–Arrhenius law

$$\tau = \tau_0 exp\left(\frac{E_a}{k_B T}\right) \quad (7)$$

where $\tau_0$ is in the $10^{-9} - 10^{-11}$ s range for SPM systems.

In the absence of an external magnetic field, the energy barrier $E_a$ can be assumed to depend on the particle volume V and the effective magnetic anisotropy $K_{eff}$ through the expression $E_a = K_{eff} V \sin^2 \theta$, where $\theta$ represents the angle between the magnetic moment of the particle and its easy magnetization axis. A linear dependence of $\ln f$ vs. $T_P^{-1}$ is expected from eq.(7) if $K_{eff}$ is assumed to be temperature-independent.

However, the extrapolation of the linear fit of the experimental data to $T^{-1} = 0$ usually gives too small, unphysical values of $\tau_0$, from $10^{-12}$ to less than $10^{-32}$ s.[51] Several attempts to fit the frequency dependence of the AC susceptibility maxima included the Vogel-Fulcher law [67] and critical slowing down [68] approaches. The sophisticated Dormann-Bessais-Fiorani model[69] of interparticle interactions tried to solve this difficulty through an interaction term in the expression of the anisotropy energy $E_a$. This attempt provided a general expression that resulted rather hard to contrast with experimental data, since it includes parameters depending on the relative location of the individual particles.

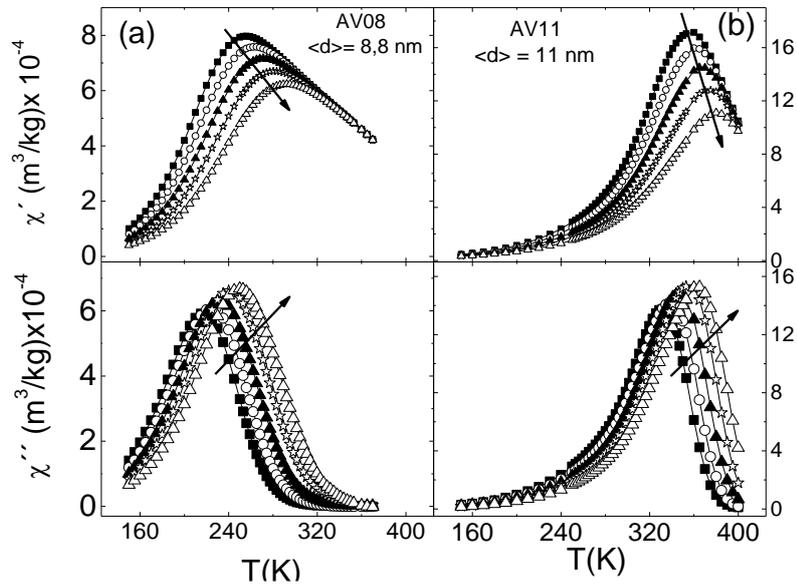

**Figure 9.** Temperature dependence of the in-phase (real) component of the magnetic susceptibility $\chi'(T)$, at different excitation frequencies for selected samples. (a) AV08 and (b) AV11. Arrows indicate increasing frequencies. Insets: Temperature dependence of the out of phase (imaginary) component, $\chi''$

Following the same approach discussed above for the temperature dependence of $H_C$, we propose to describe the $T_P(f)$ experimental data by including the explicit $K_1(T) = K_1(0)\, e^{-BT^2}$ dependence into eq.(7). By doing this a non-linear expression for $\ln \tau$ vs. $T^{-1}$ is obtained:

$$\ln \tau = \ln \tau_0 + \frac{K_1(0)V}{k_B T} exp(-BT^2) \qquad (8)$$

Figure 10 shows the good agreement between fitted curves using eq.(8) and experimental data from those samples measured within our accessible frequency range, demonstrating the suitability of the Néel–Arrhenius model to describe the magnetic relaxation. At low-temperatures, eq.(8) gives the expected linear behavior in the $\ln[\tau(T)]$ vs. $T^{-1}$ plot, whereas at high temperatures the exponential term dominates the approach to the independent $\ln\tau_0$ term, yielding realistic values of $\tau_0 \approx 10^{-10}$ s.

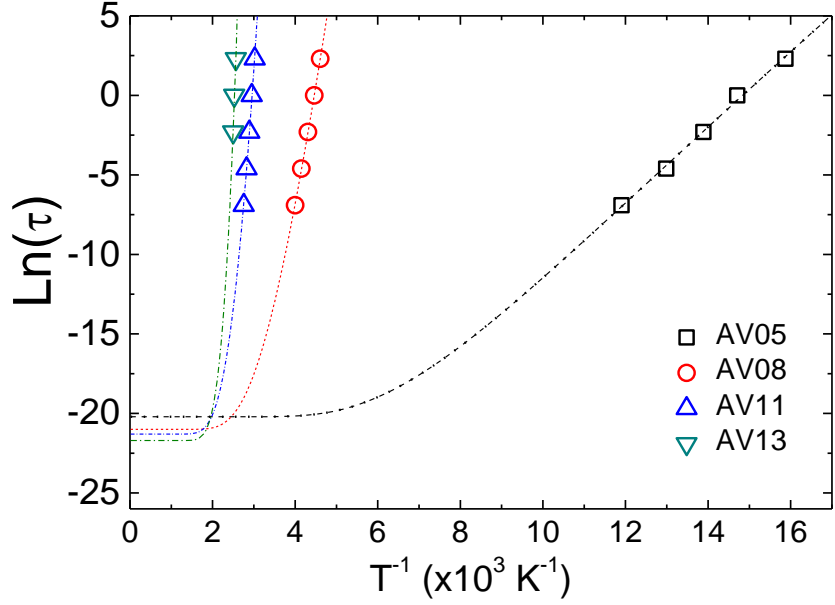

**Figure 10.** (a) Arrhenius plot of the relaxation time τ vs. $T_P^{-1}$ obtained from the imaginary component $\chi''(T)$. The lines are the corresponding fits using Eq. (7).

The $K_1(0)$ and B parameters obtained from dynamic data were found to be in agreement with the previously discussed values obtained from the fit of $H_C(T)$ curves, and consistent to those reported for bulk $CoFe_2O_4$ (see Table III). These values should be considered as the actual effective magnetic anisotropy ($K_{eff}$), since additional shape/stress contributions to the energy barrier could not be discarded. However, the close values obtained from both methods to the bulk counterpart indicate that these effects, if present, have no major influence over the overall magnetic anisotropy. Also consistent with the results from $H_C$ of the previous section, the two smallest particles AV08 and AV05 showed deviations of both $K_1(0)$ and B. Nonetheless, as our measurements of dynamic data was limited those four samples with $T_B < 400$ K, further measurements at T < 400 K would be needed to draw conclusions for the actual behavior of these parameters.

The effective magnetic anisotropy reported for many small and ultrasmall MNPs has been found to be largely enhanced with respect to the corresponding bulk materials. Furthermore, theoretical calculations have also led to expect an increase in $K_{eff}$ as the particle size decreases.[70-73] Models for this increased value have been attempted through an additional surface contribution to the total anisotropy,[74] of the form $K_{eff} = K_V + \frac{6}{d}K_S$ with $K_V$ and $K_S$ being volume and surface anisotropies for a particle of diameter $d$, although it is not clear how this approach could be applied to spherical particles, for which symmetry arguments yield a zero net contribution from the surface term. In any case, the Néel-Arrhenius or any other simple model would be expected to fail for ultra-small particles, composed by a few number of atomic layers, and a more complete approach such as the Landau-Lifschitz-Gilbert equation should be employed.[75]

## *V. Conclusions.*

Our systematic exploration of these high-anisotropy particles having <d> between 5-25 nm showed a consistent magnetic behavior over a wide range of temperatures. Interestingly, some deviations in the stoichiometry of the samples measured in macroscopic sample volumes were found to extend to the single-particle level, opening questions about the actual magnetic structure in cobalt-ferrite nanoparticles. For the smallest samples (<d> = 5 and 8 nm), non-saturating behavior of M(H) was found at 400 K and 5 K, consistent with the development of a spin-canted surface layer for decreasing particle sizes. Larger particles of the series showed some reduction of $M_S$ with respect to the bulk, pointing to the existence of partial inversion degree. Furthermore, our systematic measurements of the static and dynamic magnetic properties in the series of $Co_xFe_{3-x}O_4$ nanoparticles provided an experimental framework to check the validity of the Néel-Arrhenius model for single-domain nanoparticles. The systematic analysis of the thermal dependence of coercive field for different particle sizes showed that the deviations, usually reported in high-anisotropy MNPs, from the Néel-Arrhenius magnetic relaxation model can be accounted for by considering the temperature dependence of the *$K_1(T)$* in the fit of the experimental data. The same straightforward approach of including the thermal variation of *$K_{eff}$* explained the magnetic dynamics of our nanoparticles as obtained from ac susceptibility measurements. Indeed, making use of an empirical expression for *$K_1$*(T) in bulk materials we were able not only to fit the frequency dependence of the ac susceptibility peaks but to obtain values of the characteristic response time $τ_0$ more realistic than those usually reported in the literature. Our approach demonstrates that it is possible to analyze the temperature dependence of the magnetic parameters of high-anisotropy MNPs without the need of artificial corrections to the Neel–Arrhenius relaxation framework, which correctly describes the dynamic response of single-domain magnetic nanoparticles.

*Acknowledgements*

This work was supported by the Spanish Ministerio de Economia y Competitividad (MINECO, project MAT2010-19326 and MAT2013-42551). The authors are indebted to Dr. A. Goméz Roca and Dr. M.P. Morales for their advice on sample preparation. GFG wish to thank Professor R.F. Jardim for valuable discussions that stimulated this investigation. Accessibility to LMA-INA and SAI-UZ facilities are also acknowledged.